\documentclass[aps,prl,floatfix,showpacs,twocolumn]{revtex4}
\usepackage{amsmath,amssymb}
\usepackage{epsfig}

\begin{document}

%\twocolumn[
\hsize\textwidth\columnwidth\hsize\csname@twocolumnfalse\endcsname

\title{Spin Accumulation in the Extrinsic Spin Hall Effect}

\author{Wang-Kong Tse$^{1}$, J. Fabian$^{2}$, I. \v{Z}uti\'c$^{1,3}$, and S. Das Sarma$^{1}$}

\affiliation{$^1$Condensed Matter Theory Center, Department of Physics,
University of Maryland at College Park, College Park, Maryland 20742-4111, USA \\
$^2$Institute for Theoretical Physics, University of Regensburg, 
93040 Regensburg, Germany \\
$^3$Center for Computational Materials Science, Naval Research
Laboratory, Washington, D.C. 20735, USA}

\begin{abstract}
The drift-diffusion formalism for spin-polarized carrier transport 
in semiconductors is generalized
to include spin-orbit coupling. The theory is applied to treat
the extrinsic spin Hall effect using realistic boundary conditions.
It is shown that carrier and spin diffusion lengths are modified by
the presence of spin-orbit coupling and that spin accumulation due
to the extrinsic spin Hall effect is strongly and qualitatively influenced by boundary
conditions. Analytical formulas for the spin-dependent carrier 
recombination rates and inhomogeneous spin densities and currents 
are presented.
\end{abstract}
%]

\maketitle
\newpage

In the presence of spin-orbit coupling, either due to impurities
or due to host lattice ions, carriers of opposite spins tend to scatter 
into opposite directions. With an electric field induced
(longitudinal) motion under bias, the spin-orbit scattering results in a transverse 
spin current and spin accumulation, as first predicted by D'yakonov and 
Perel'~\cite{Dyak1,Dyak2}, and later revisited by others ~\cite{Hirsch}.
This effect, which is now called the extrinsic spin Hall effect (SHE)
~\cite{extrinsic}, has been recently demonstrated experimentally 
in n-GaAs and n-InGaAs thin films~\cite{Aws} and in two-dimensional
electron gas confined within (110) AlGaAs quantum wells~\cite{Sih}. 
The signature of the effect is opposite spin accumulation at
the edges of the sample, with spin polarization perpendicular to the
transport plane.

This paper has two goals. First, we present a formalism for carrier drift
and diffusion in inhomogeneous spin-polarized semiconductors in the presence 
of spin-orbit coupling and spin-dependent band-to-band electron-hole 
recombination. The formalism, which is a generalization of a previous spin
and charge drift-diffusion theory \cite{Zutic1,Zutic2}, 
applies to both unipolar and bipolar cases, 
the former being a subclass of the latter. Second, we apply
the formalism to explain the main qualitative features of  
spin accumulation in the extrinsic SHE in the optical orientation experiment, 
for two different boundary conditions: (a) uniform generation of electron-hole
pairs, and (b) edge generation of nonequilibrium electrons. In both 
cases spin accumulation throughout the sample is calculated analytically.
We find that spin-orbit interaction modifies the carrier
and spin diffusion lengths and that the spin accumulation profile
depends, qualitatively, on the specific boundary conditions,
implying that interpretation of extrinsic SHE requires detailed case-by-case
considerations for specific experimental and sample geometries.

Consider spin-polarized transport in an inhomogeneous nonmagnetic semiconductor 
in the presence of electric field ${\boldsymbol E}$. If spin-orbit
coupling is present, causing skew scattering and side jump, the phenomenological 
expression for the carrier ($c=n$ for electrons
and $c=p$ for holes) charge current density in the $i^{\textrm{th}}$ direction is readily obtained by generalizing the
Dyakonov-Perel' prediction \cite{Dyak1,Dyak2}:
\begin{eqnarray}
J_{c\lambda,i} = &&q\mu_{c\lambda}c_{\lambda}E_i\pm
qD_{c\lambda}{\partial_i}c_{\lambda}+q\lambda\nu_{c\lambda}c_{\lambda}\epsilon_{ijz}E_j
\nonumber \\
&&+q\lambda\delta_{c\lambda}\epsilon_{ijz}
{\partial_j}c_{\lambda}.
\label{sh1}
\end{eqnarray}
Here the upper (lower) sign is for electrons (holes) and $\lambda$
is the spin index; $q$ is the proton charge.  The first two terms are 
conventional (longitudinal) carrier
drift and diffusion, respectively, with $\mu$ and $D$ denoting the
spin-dependent mobility and diffusivity. The third (fourth) term represents 
the effect of skew spin-orbit scattering and side jump on drift (diffusion). The 
effects of the scattering are in the transverse direction to $\boldsymbol E$ and
are opposite for spin up and down carriers. (Holes are treated here as
spin doublets, which is appropriate in low-dimensional 
structures with heavy and light hole band splitting; otherwise hole
spin does not matter, as we will argue below). The corresponding transport 
parameters are transverse mobility $\nu$ and transverse diffusivity $\delta$;
they are proportional to the spin-orbit coupling strength.

It is more illuminating to introduce the charge, $J_c =
J_{\uparrow}+J_{\downarrow}$, and spin, 
$J_s = J_{\uparrow}-J_{\downarrow}$, currents. 
In terms of carrier ($c=c_{\uparrow}+c_{\downarrow}$) and spin 
($s_c = c_{\uparrow}-c_{\downarrow}$) densities, the currents
are 
\begin{eqnarray}  
&J_{c,i}& = q(\mu_c c+\mu_{sc} s_c)E_i \pm
q(D_c{\partial_i}c+D_{sc}{\partial_i}s_c) \nonumber \\
&&+q\epsilon_{ijz}E_j(\nu_c s_c+\nu_{sc}
c)+q\epsilon_{ijz}(\delta_c{\partial_j}s_c+\delta_{sc}{\partial_j}c),
 \label{sh2} \\
%iz\\
%&J_{c,i}& = q(\mu_c c+\mu_{sc} s_c)E_i \pm
%q(D_c{\partial_i}c+D_{sc}{\partial_i}s_c) \nonumber \\
%&&+q\epsilon_{ijz}E_j(\nu_c s_c+\nu_{sc}
%c)+q\epsilon_{ijz}(\delta_c{\partial_j}s_c+\delta_{sc}{\partial_j}c),
%\nonumber \\ \label{sh2}
%\\
&J_{sc,i}& = q(\mu_{sc} c+\mu_c s_c)E_i \pm
q(D_c{\partial_i}s_c+D_{sc}{\partial_i}c) \nonumber \\ 
&&+q\epsilon_{ijz}E_j(\nu_c c+\nu_{sc}
s_c)+q\epsilon_{ijz}(\delta_c{\partial_j}c+\delta_{sc}{\partial_j}s_c)
 \label{sh3}.
\end{eqnarray} 
The transverse carrier charge and spin mobilities are given respectively by
$\nu_c = (\nu_{c\uparrow}+\nu_{c\downarrow})/2$ and
$\nu_{sc} = (\mu_{c\uparrow}-\nu_{c\downarrow})/2$, while the
transverse carrier charge and spin diffusivities are
$\delta_c = (\delta_{c\uparrow}+\delta_{c\downarrow})/2$ and
$\delta_{sc} = (\delta_{c\uparrow}-\delta_{c\downarrow})/2$. The
corresponding longitudinal quantities are defined similarly.

Eqs.~(\ref{sh2}) and (\ref{sh3}) succinctly describe the appearance 
of the transverse spin drift and diffusion in the presence of 
longitudinal charge transport, which is the essence of the extrinsic SHE. 
If the longitudinal current is spin-polarized, 
the above equations describe the anomalous
Hall effect~\cite{Karplus} and the appearance of the transverse Hall voltage. 

To further develop the formalism, we need to include electron-hole
recombination and spin relaxation. In the presence of spin-orbit
coupling, spin-dependent selection rules \cite{opo} for 
band-to-band transitions need to be considered. In general, the 
continuity equation reads:
\begin{eqnarray}
\pm{\partial_i}J_{c\lambda,i}/q & = & 
 B_1\left (c_{\lambda}\bar{c}_{\lambda} - 
c_{0\lambda}\bar{c}_{0\lambda}\right ) \nonumber \\
&+& B_2\left (c_{\lambda}\bar{c}_{\bar{\lambda}} - 
c_{0\lambda}\bar{c}_{0\bar{\lambda}}  \right ) 
+  \left (c_{\lambda}-c_{\bar{\lambda}} \right)/2T_{1c}. 
%\mp\frac{1}{q}{\partial_i}J_{c\lambda,i} & = & 
% B_1\left (c_{\lambda}\bar{c}_{\lambda} - c_{0\lambda}\bar{c}_{0\lambda}\right )
%+ B_2\left (c_{\lambda}\bar{c}_{\bar{\lambda}} - 
%c_{0\lambda}\bar{c}_{0\bar{\lambda}}  \right ) \nonumber
%\\
%& + & \frac{c_{\lambda}-c_{\bar{\lambda}}}{2T_{1c}}.
\label{sh4}
\end{eqnarray}
Here $\bar{c}$ is $p(n)$ if $c$ is $n(p)$, $c_0$ is the equilibrium
carrier density, and $T_{1c}$ is the $T_1$ time for spin flipping.

The spin-preserving recombination rate coefficient $B_1$, 
as well as the spin-flip coefficient $B_2$, can be calculated
by generalizing the unpolarized case \cite{rec1,rec2}. 
The valence band of zinc-blende semiconductors consists
of three subbands: heavy-hole, light hole, and split-off hole 
bands. We neglect the split-off band as 
the energy splitting $\Delta \gg k_B T$ at temperatures $T$
lower than or around room temperature. By explicitly taking into account
the angular momentum of the heavy hole and light hole states,
and using the optical selection rules for the states \cite{opo}, 
we arrive at the following expressions for spin-conserving and
spin flip recombination constants:
\begin{eqnarray}
B_1 & = & C \frac{\lbrack
m_h/(m_c+m_h) \rbrack^{3/2}+\frac{2}{3}\lbrack
m_l/(m_c+m_l) \rbrack^{3/2}}{m_h^{3/2}+m_l^{3/2}},
\label{rec5} \\
B_2 & = & \frac{1}{3} C\frac{\lbrack m_h/(m_c+m_h) 
\rbrack^{3/2}}{m_h^{3/2}+m_l^{3/2}},
\label{rec6}
\end{eqnarray}
where $C$ depends on the ``maximum'' electron energy $\hbar\omega_{\rm max}
= \epsilon_g+k_B T/2$ ($\epsilon_g$ is the energy gap) and temperature:
\begin{eqnarray}
C(\hbar\omega_{\rm max}, T) = \frac{4 e^2}{\hbar^2 m^2
c^3}\left(\frac{2\pi\hbar^2}{k_B T}\right)^{\frac{3}{2}}P^2 n_r
\hbar\omega.
\label{rec7}
\end{eqnarray}
Here $P$ is the momentum matrix
element for optical transitions, $n_r$ is the refractive index, 
$m_c$ ($m_h$ and $m_l$) is the band mass of electrons 
(heavy holes and light holes).

Eqs.~(\ref{sh1}) and (\ref{sh4}), together with Poisson's equation form a
closed set of nonlinear equations whose solution determines charge and spin
densities and currents in a semiconductor with spin-orbit scattering
included. In general, these equations need to be 
solved numerically for specific cases of interest. Our next goal is
to introduce qualitative features of spin accumulation in two
cases of experimental interest that allow analytical solutions
and form a starting point to discuss the concepts and issues 
to be encountered in more complex situations involving SHE and spin-orbit coupling effects 
in transport. We consider a p-type semiconductor with nondegenerate electron (minority)
density induced optically. The resulting spin accumulation via extrinsic SHE
can be deduced in a manner similar to spin orientation experiments. 
We assume that the injected electron density is well below the donor 
density. Further simplification follows from the fact that mobilities and 
diffusivities are spin independent in the nondegenerate regime \cite{Zutic1}. 
Finally, we assume unpolarized holes since hole spin relaxation in zinc-blende
semiconductors is extremely fast \cite{Zutic3}. 
The only carriers of interest are then spin-polarized electrons.

%iz 
%Using the above assumptions, Eqs.~(\ref{sh2}), (\ref{sh3}), and (\ref{sh4}), 
Using the above assumptions, Eqs.~(\ref{sh2})--(\ref{sh4}), 
give the drift-diffusion equations for electron spin and carrier density:
\begin{eqnarray}
\nabla^2 s+\frac{q}{k_B
T}\left(s\nabla\cdot\boldsymbol{E}+\boldsymbol{E}\cdot\nabla s\right)
+\frac{\zeta q}{k_B T}\left(\nabla n\times\boldsymbol{E}\right)_z
&& \nonumber \\
+\left(wp+1/T_{1n}\right)s = 0,&& \label{sh6}
\end{eqnarray}
\begin{eqnarray}
\nabla^2 n+\frac{q}{k_B
T}\left(n\nabla\cdot\boldsymbol{E}+\boldsymbol{E}\cdot\nabla
n\right)+\frac{\zeta q}{k_B T}\left(\nabla s\times\boldsymbol{E}\right)_z
&& \nonumber \\
+w\left(np-{n_0}{p_0}\right)=0,&& \label{sh7}
\end{eqnarray}
where $w = (B_1+B_2)/2$ and $\zeta=\nu_n/\mu_n = \delta_n/D_n$  
characterizes the spin-orbit coupling strength. The spin quantization
axis is taken to be $z$.  We take ${\boldsymbol E} = E\hat{\boldsymbol{y}}$
and consider $x$ to be the transverse direction, the slab boundaries 
being at $x=0$ and $x=a$, so that all the quantities of interest will have $x$ dependence only. 
Denoting the electron recombination time as $\tau_n = 1/(wN_a)$, where $N_a$ is the acceptor
density, and spin relaxation time as $\tau_s = 1/(\tau_n^{-1}+T_{1n}^{-1})$, 
Eqs.~(\ref{sh6}) and (\ref{sh7}) become
\begin{eqnarray}
d^2s/dx^2+(\zeta qE/k_B T) dn/dx-s/L_s^2& =& 0, \label{sh8} 
%\frac{d^2s}{dx^2}+\frac{\zeta qE}{k_B
%T}\frac{dn}{dx}-\frac{s}{L_s^2}& =& 0, \label{sh8} 
\\
d^2 n/dx^2+ (\zeta
qE/k_B T){ds}/{dx}-(n-n_0)/L_n^2& = &0, \label{sh9}
\end{eqnarray}
where the longitudinal spin and charge diffusion lengths
are defined as $L_s = \sqrt{D_s\tau_s}$ and $L_n =
\sqrt{D_n\tau_n}$, respectively. In deriving the above equations
we have neglected terms of order $s^2$, $s\delta n$, and $\delta n^2$
relative to $\delta n=n-n_0$ \cite{Smith}. Spin-charge coupling 
in Eqs.~(\ref{sh8}) and (\ref{sh9}) is apparent through the derivatives of
spin and charge densities. 

\begin{figure}
\includegraphics[width=5.9cm,angle=270]{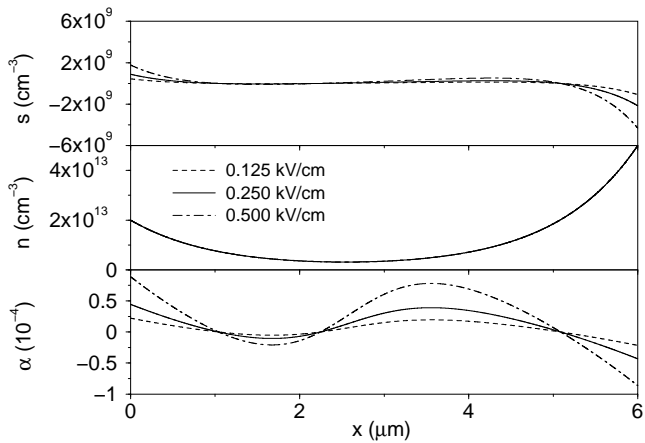}
\caption{Calculated spin density $s$, electron density $n$, and
polarization $\alpha$ for BC II. We have used $\zeta = 10^{-4}$ and
applied electric field strengths $0.125$ kV/cm, $0.250$ kV/cm, and $0.500$ kV/cm. 
While a signature of the extrinsic SHE is the opposite spin accumulation at the edges of 
the sample, spin accumulation is large also inside.}
\label{fig:snp0}
\end{figure}

We first solve the transverse spin and carrier diffusion for the case of  a 
uniformly illuminated slab (boundary condition BC I), with electron-hole spin-unpolarized
generation rate $G$. We assume that carrier recombination at the edges is
not significant so that it is reasonable to impose a uniform (here zero) 
electron transverse current: $J_{nx}(x)=0$. As for spin current, we take 
$J_s(x=0)=J_s(x=a)=0$, implying that spin-flip scattering at the edges is
moderate. The solution to the drift-diffusion equation (\ref{sh8}) in the presence of
spin-orbit scattering is 
\begin{eqnarray}
s(x) = \zeta \frac{qEL_s}{k_BT} G\tau_n {\rm sech}(a/2L_s)\sinh(x/L - a/2L).
\end{eqnarray}
Spin polarization profile is given by $\alpha(x) = s(x)/G\tau_n$, since
$G\tau_n \gg n_0$ is the average electron density generated in steady-state conditions
under illumination. This simple solution demonstrates the essential physics
behind spin accumulation in extrinsic SHE: (i) Spin accumulation
increases linearly with $E$, with possible slight electric field modulation due
to the dependence of $L_s = L_s(E)$ (not discussed here); 
(ii) The magnitude of spin polarization is
proportional to the strength of the spin-orbit scattering as well
as to the ratio of the voltage drop over $\min(L_s,a)$ and thermal energy;
(iii) For $a\ll L_s$ the accumulation at the edges is linearly proportional to $a$, 
while for $a \gg L_s$, the accumulation is independent of $a$. Finally,
(iv), while $\alpha(x) \sim x$ for $a \ll L_s$,  spin accumulation is significant
only within the spin diffusion length from the edges when $a \gg L_s$.

A question now arises: How universal (i.e. independent of boundary conditions) 
is the qualitative behaviour discussed above? To answer this question we introduce
different boundary conditions (BC II), describing the physics of
carrier injection at $x=0$ and $x=a$, while assuming vanishing spin currents at the edges: 
$n(x=0) = n_1, n(x=a) = n_2;\; J_{s,x}(x=0) = J_{s,x}(x=a)=0$.
These conditions mimic the case of a p-doped
base in a pnp spin-polarized transistor \cite{fabian}, where the
electron injection level can be controlled by the biases to
the emitter and collector. In the case discussed here the longitudinal
current would flow perpendicular to the transistor current, which 
would thus be spin-polarized due to the extrinsic SHE. 
The vanishing boundary conditions for $J_{s,x}$ 
reduce to
\begin{equation}
\frac{ds}{dx}\left\vert_{x=0}=-\frac{\zeta qE}{k_B T}n_1\right., \quad 
\frac{dn}{dx}\left\vert_{x=a}=-\frac{\zeta qE}{k_B T}n_2\right. .
\label{sh12}
\end{equation}
Solving Eqs.~(\ref{sh8}), (\ref{sh9}) with the above boundary conditions gives
the spin and electron densities inside the slab:
\begin{eqnarray}
&&s(x)=\frac{\zeta qE}{k_B T(L_1^{-2}-L_2^{-2})}
\nonumber
\\
&&\left\{\frac{1}{L_1}\mathrm{cosech}(\frac{a}{L_1})\left[n_1 
\mathrm{cosh}(\frac{a-x}{L_1})-n_2 \mathrm{cosh}(\frac{x}{L_1})\right]\right.
\nonumber
\\
&&\left.-\frac{1}{L_2}\mathrm{cosech}(\frac{a}{L_2})\left[n_1 \mathrm{cosh}(\frac{a-x}{L_2})
-n_2 \mathrm{cosh}(\frac{x}{L_2})\right]\right\},
\nonumber
\\
\label{sh14}
\end{eqnarray}
and
\begin{eqnarray}
&&n(x)-n_0=\frac{1}{(L_1^{-2}-L_2^{-2})}
\nonumber
\\
&&\left\{(\frac{1}{L_1^2}-\frac{1}{L_s^2})\mathrm{cosech}(\frac{a}{L_1})\left[n_1 
\mathrm{sinh}(\frac{a-x}{L_1})+n_2
\mathrm{sinh}(\frac{x}{L_1})\right]\right.
\nonumber
\\
&&\left.-(\frac{1}{L_2^2}-\frac{1}{L_s^2})\mathrm{cosech}(\frac{a}{L_2})\left[n_1 
\mathrm{sinh}(\frac{a-x}{L_2})+n_2 \mathrm{sinh}
(\frac{x}{L_2})\right]\right\}.
\nonumber
\\
\label{sh15}
\end{eqnarray}
Here we introduce new transverse spin diffusion lengths, $L_1$ and $L_2$:
\begin{equation}
L_{1,2}^{-2}=\gamma\pm\sqrt{\gamma^2-\left(1/L_s L_n\right)^2},
\label{sh16}
\end{equation}
where
\begin{equation}
\gamma=({1}/{2})\left[{1}/{L_s^2}+{1}/{L_n^2}+\left({\zeta qE}/{k_B
  T}\right)^2\right]. \label{sh17}
\end{equation}
There is a critical value of the field which separates 
the regimes of strong and weak spin-charge 
coupling---Eqs.~(\ref{sh8}) and (\ref{sh9}) reduce to the ordinary spin
and charge diffusion equations when $E \ll E_s, E_n$, where $E_s = k_B
T/(\zeta q L_s)$ and $E_n = k_B T/(\zeta q L_n)$ are the values of
the critical fields with respect to spin and charge diffusion. In this case 
$\gamma \simeq (L_s^{-2}+L_n^{-2})/2$ and $L_{1,2} \simeq L_{s,n}$.
When $E \gtrsim E_s, E_n$, $\gamma$ becomes dependent on the electric field
and it is in this regime the spin-field relation deviates from linearity. 

For quantitative understanding we take our semiconductor to be 
GaAs  at room temperature \cite{300K}, with doping density
$N_a = 3 \times 10^{15}$ cm$^{-3}$, and transverse size of
$a = 6$ $\mu$m (which is much greater than $L_s$). The spin-orbit
scattering strength is taken to be $\zeta=10^{-4}$, reflecting weak spin-orbit
coupling in GaAs. Finally, the boundary conditions
for electron density are $n(0)=2\times 10^{13}$/cm$^3$ and 
$n(a)=5\times 10^{13}$/cm$^3$.
Figure~\ref{fig:snp0} shows the profiles of spin and electron densities, as
well as spin polarization $\alpha=s/n$, for several values of $E$.
In contrast to the purely diffusive behavior exhibited by $n$,
spin density along the slab is weakly oscillatory. Both spin density and spin polarization 
attain maximum magnitudes at the edges. What is interesting is that, unlike
in BC I [conclusion (iv)], spin accumulation
here is significant throughout the sample, not only within $L_s$ from the edges.
(We find that spin polarization is enhanced by a decade when considering
the extrinsic SHE at 77 K.)

\begin{figure}
\includegraphics[width=5.3cm,angle=270]{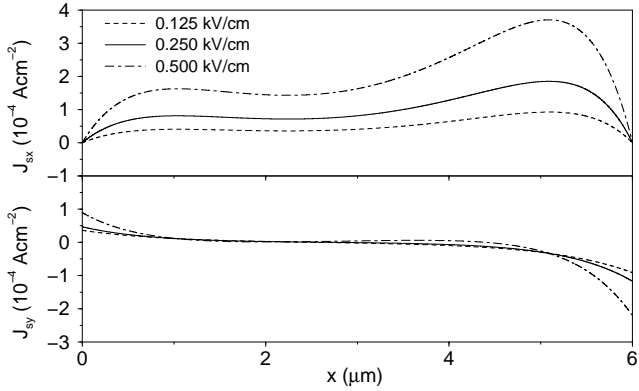}
\caption{\label{fig:Jsnp0} Calculated spin current $J_s$ in $x$ (top) 
and $y$ (bottom) directions for BC II and parameters as in Fig.~\ref{fig:snp0}.}
\label{fig:jsnp0}
\end{figure}

Spin current is not conserved and flows inside the sample, being restricted
to zero by our choice of boundary conditions. From Eq.~(\ref{sh3}) we obtain
the spin current density in the $x$ direction:
\begin{eqnarray}
&&J_{s,x}(x)=-\frac{q \nu_n E}{(L_1^{-2}-L_2^{-2})L_s^2}
\nonumber
\\
&&\left\{\mathrm{cosech}\frac{a}{L_1})\left[n_1 \mathrm{sinh}(\frac{a-x}{L_1})+n_2 \mathrm{sinh}(\frac{x}{L_1})\right]\right.
\nonumber
\\
&&\left.-\mathrm{cosech}(\frac{a}{L_2})\left[n_1 \mathrm{sinh}(\frac{a-x}{L_2})
+n_2 \mathrm{sinh}(\frac{x}{L_2})\right]\right\},
\nonumber
\\
\label{sh18}
\end{eqnarray}
and in the $y$ direction:
\begin{eqnarray}
&&J_{s,y}(x)=\frac{1}{(L_1^{-2}-L_2^{-2})}
\nonumber
\\
&&\left\{\left[\frac{q\nu_n E^2}{k_B T}+\delta_n(\frac{1}{L_1^2}-\frac{1}{L_s^2})\right]
\frac{1}{L_1}\mathrm{cosech}(\frac{a}{L_1})\right. 
\nonumber
\\
&&\left[n_1 \mathrm{cosh}(\frac{a-x}{L_1})-n_2 \mathrm{cosh}(\frac{x}{L_1})\right]
\nonumber
\\
&&-\left[\frac{q\nu_n E^2}{k_B T}+\delta_n(\frac{1}{L_2^2}-\frac{1}{L_s^2})\right]
\frac{1}{L_2}\mathrm{cosech}(\frac{a}{L_2})
\nonumber
\\
&&\left.\left[n_1 \mathrm{cosh}(\frac{a-x}{L_2})-n_2 \mathrm{cosh}(\frac{x}{L_2})\right]\right\}.
\nonumber
\\
\label{sh19}
\end{eqnarray}
Spin current in the $x$ direction flows in the direction of the spin gradient, while
the $y$ component changes sign inside the slab, reflecting the fact that
spin-up and spin-down electrons are deflected into opposite directions
due to spin-orbit scattering by impurities. 

\begin{figure}
\includegraphics[width=5.5cm,angle=270]{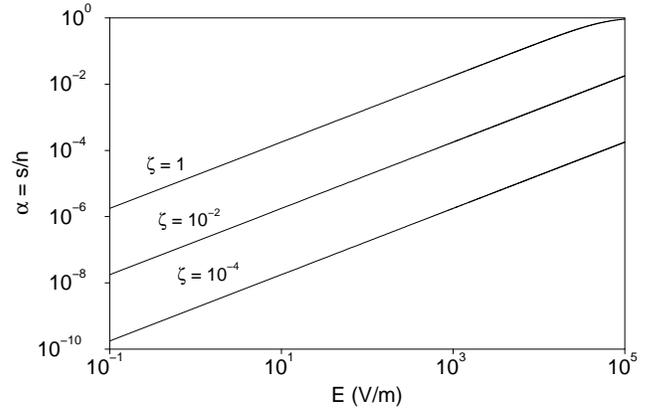}
\caption{\label{fig:alphas0Ep} Calculated spin polarization $\alpha$ at $x =
0$ as a function of electric field $E$ for BC II and $\zeta = 10^{-4}$,  $10^{-2}$, and $1$.
Only in the last (unphysical) case $\alpha$ starts to saturate
for large $E$, as the dependence of $L_1$ and $L_2$ on $E$ sets in.}
\label{fig:alpha}
\end{figure}

Finally, we wish to see at which electric field spin and carrier diffusion lengths
$L_{1,2}$ are modified and induce nonlinear behavior in $\alpha$. 
Figure~\ref{fig:alpha} shows $\alpha(E)$ for different $\zeta$. Spin
polarization varies linearly with $E$, except at electric fields as large as $10^5$ kV/cm 
and spin-orbit couplings $\zeta\approx 1$, a clearly unphysical case considered here only
to illustrate the scope of linear behavior.

In conclusion, we have presented a drift-diffusion formalism which takes into account
spin-orbit scattering and spin-dependent carrier recombination. We have calculated spin accumulation
in the extrinsic spin Hall regime and introduced spin-orbit dependent carrier and spin diffusion 
lengths. We have found that spin accumulation is strongly influenced by boundary conditions
and thus by specific spintronic device design. Our theory is applicable to any complex
device setting in which extrinsic spin Hall effect is expected to play a role. We expect 
similar strong dependence of the ``intrinsic'' SHE on boundary conditions too, making it difficult
to distinguish intrinsic and extrinsic SHE in general.

This work was supported by the US ONR, NSF,
and National Research Council (I.\v{Z}.). 
 
%\bibliographystyle{apsrev}
%\bibliography{spintronics}

\begin{thebibliography}{13}
\bibitem{300K} For $T = 300$ K, the transport parameters for the minority 
electron are taken as~\protect{\cite{Zutic1}}: 
diffusivity $D_n = 103.6$ cm$^2\cdot$s$^{-1}$, mobility $\mu_n = 4000$
cm$^2\cdot$V$^{-1}\cdot$s$^{-1}$, recombination rate $w = (1/3) \times
10^{-5}$ cm$^3\cdot$s$^{-1}$ and spin-flip time $T_1 = 0.2$ ns. From these
the calculated values of electron-hole recombination and spin diffusion lengths
are, respectively, $L_n = 1$ $\mu$m and $L_s = 0.8$ $\mu$m.  
The intrinsic concentration is $n_i = 1.8 \times 10^6$ cm$^{-3}$.
\end{thebibliography}

\end{document}